\documentstyle[prb,aps,epsfig,floats,goodfloat]{revtex}

\newcommand{\bc}{\begin{center}}
\newcommand{\ec}{\end{center}}
\newcommand{\be}{\begin{equation}}
\newcommand{\ee}{\end{equation}}
\newcommand{\beqn}{\begin{eqnarray}}
\newcommand{\eeqn}{\end{eqnarray}}
\newcommand{\nsw}{N_{\mathrm{sweep}}}
\newcommand{\nsa}{N_{\mathrm{samp}}}

\begin{document}
\draft

\twocolumn[\hsize\textwidth\columnwidth\hsize\csname@twocolumnfalse%
\endcsname

\title{
The $\pm J$ Spin Glass: Effects of Ground State Degeneracy
}

\author{Matteo Palassini}
\address{
Department of Pharmaceutical Chemistry, University of California,
San Francisco, CA 94118
}
\author{A. P. Young}
\address{Department of Physics, University of California, Santa Cruz, 
CA 95064}

\date{\today}

\maketitle

\begin{abstract}
We perform Monte Carlo simulations of the Ising spin glass at low temperature
in three dimensions with a $\pm J$ distribution of couplings. Our results
display crossover scaling between $T=0$ behavior, where the order parameter
distribution $P(q)$ becomes trivial for $L \to \infty$, and finite-$T$ behavior,
where the non-trivial part of $P(q)$ has a much weaker dependence on $L$,  and
is possibly size independent.
\end{abstract}

]

\section{Introduction}
Several papers\cite{km,py,kpy,mp}
have recently studied the Ising spin glass in three dimensions with a
Gaussian distribution of bonds at low and zero temperature. {\em From data
obtained on small sizes}\/ these papers deduce that the order parameter
distribution function, $P(q)$, is non-trivial at finite-$T$, i.e. in addition to
two peaks,
symmetric about $q=0$, there is also a continuous part between the peaks whose
weight does not decrease with size\cite{extrap}. 
This indicates the existence of a nontrivial energy landscape, 
i.e. of macroscopic excitations, involving a finite fraction of the system,
that cost a finite energy in the thermodynamic limit.
This aspect of the results is
consistent with the replica symmetry breaking picture of Parisi\cite{parisi}.
By contrast, the droplet theory\cite{fh,bm} predicts that the weight in the
continuous part of the distribution should vanish
like $L^{-\theta}$
as the (linear) size of the
system $L$ increases, where $\theta$ is a positive exponent.
In both theories, because the ground state is unique
(apart from inverting all the 
spins), it follows that the weight in the ``tail'' of the distribution
tends to zero (proportional to $T$) as $T \to 0$
and the positions of the peaks tend to
$\pm 1$. The purpose of this paper is to see how these results are modified
for a spin glass with a bimodal distribution (also called the $\pm J$
distribution), where the interactions have values $\pm 1$, where
there is a large ground state degeneracy and a finite ground state
entropy per spin.

One might possibly imagine that, since the system with the $\pm J$
distribution has a finite ground state entropy, its behavior at zero
temperature would be similar to that of a model with continuous distribution
at finite-$T$. If this were true then, according to the numerical
results,\cite{km,py,kpy,mp} $P(q)$ would be non-trivial at $T=0$ whereas
according to the droplet theory $P(q)$ would be trivial. 

However, this notion has been contested by Krzakala and 
Martin\cite{km2} (referred to henceforth as KM)
who argue that entropy effects cause one ``valley'' in the
$T=0$ energy landscape of the $\pm J$ model to dominate and consequently the
weight in the tail vanishes like $ L^{-\lambda}$, where $\lambda$ is a
positive exponent (discussed below), even if 
the energy landscape is non-trivial.
At finite-$T$, KM argue that the weight is finite for large $L$, so, by
implication, there must be a crossover at some scale $L_c(T)$ from the
$L^{-\lambda}$ behavior for $L < L_c(T)$ to a value independent of $L$ at
larger sizes. One can also generalize the KM argument to the droplet
model, in which case there is still a crossover, between $L^{-(\lambda+\theta)}$
behavior at smaller $L$ and $L^{-\theta}$ behavior at larger $L$.
Overall, in the KM scenario, the only difference between the
continuous and the $\pm J$ distributions for $L \gg L_c(T)$
is that the position of
the peaks in $P(q)$ are different for $T \to 0$. Denoting the peak positions
by $\pm q_0$, then one has $q_0 <1$
for the $\pm J$ distribution whereas $q_0$ = 1 for a
continuous distribution.

Here we display, we believe for the first time, the crossover between $T=0$
and finite-$T$ behaviors. Further motivation for our work is to clarify 
conflicting results for ground state properties.
Berg et al.\cite{berg} used a multicanonical Monte Carlo technique to determine
$P(q)$ at $T=0$ finding results consistent with trivial behavior with $\lambda
= 0.72 \pm 0.12$ (but also not ruling out the possibility of nontrivial
behavior). Hartmann\cite{hart1} used a genetic optimization algorithm
finding initially a nontrivial $P(q)$, but the results 
were biased\cite{sandvik} because the degenerate ground states were not 
sampled with equal probability. Subsequently Hartmann\cite{hart2} 
developed an improved method and found a trivial $P(q)$ with 
$\lambda = 1.25 \pm 0.05$, and suggested that this
supports the droplet picture. 
Very recently Hatano and
Gubernatis\cite{hg} (referred to as HG) have performed a ``bi-variate
multi-canonical'' Monte Carlo study, finding that $P(0)$ drops dramatically
at low-$T$ as $L$ increases. Though they do not extract the exponent
$\lambda$, from the figures in their paper, it appears that $\lambda$ is
significantly larger than Hartmann's value. They too argue that their results
provide evidence for the droplet picture.
However, Marinari et al.\cite{mprz}
have recently claimed, on the basis of their own
simulations, that the results of HG are not equilibrated and
their conclusions are therefore invalid. 
Finally, recent work\cite{hhsd} finds a
nontrivial energy landscape and also, apparently, a
{\em nontrivial} $P(q)$ at $T=0$.
It therefore seems useful to try to decide between these different results.
Our data at the lowest temperatures imply a trivial $P(q)$ at $T=0$ and our
estimate for $\lambda$ is
consistent with that of Berg et al.\cite{berg} but not with that
of Hartmann\cite{hart2} or HG.

The Hamiltonian is given by
\begin{equation}
{\cal H} = -\sum_{\langle i,j \rangle} J_{ij} S_i S_j ,
\label{ham}
\end{equation}
where the sites $i$ lie on a 
simple cubic lattice in dimension $d=3$ with $N=L^3$ sites 
($L \le 10$), $S_i=\pm
1$, and the $J_{ij}$ are nearest-neighbor interactions taking values $\pm 1$
with equal probability. We do not apply the constraint 
$\sum_{\langle i,j \rangle} J_{ij} = 0$, which is imposed in some related
work.
However, we expect that the crossover
from $T=0$ to finite-$T$ behavior will be similar in the two models.
Periodic
boundary conditions are applied. 
We focus on the distribution of the spin overlap, $q$, where
\begin{equation}
q = {1 \over N} \sum_{i=1}^N S_i^{(1)} S_i^{(2)} ,
\label{q}
\end{equation}
in which ``$(1)$'' and ``$(2)$'' refer to two independent copies (replicas)
of the system with identical
bonds.

\begin{figure}
\begin{center}
\epsfxsize=\columnwidth
\epsfbox{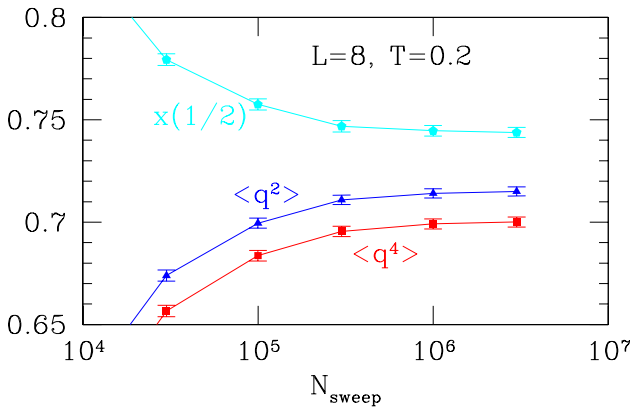}
\end{center}
\caption{
An equilibration plot for $L=8, T = 0.20$, for the second and fourth
moment of $P(q)$, and for $x(1/2)$, the average of $P(q)$ over the
interval $|q|\le 1/2$. For better viewing, the
data for $\langle q^4 \rangle$ and $x(1/2)$ have been shifted upwards by 
0.14 and 0.67, respectively. For each value of $N_{sweep}$, the averages
were measured over the last $N_{sweep}/3$ MC sweeps.}
\label{equil}
\end{figure}

Simulations of spin glasses at very low temperatures are now possible, at
least for modest sizes, using the parallel
tempering Monte Carlo method\cite{huk_nem,marinari_buda}, where
one simulates replicas of the system
at $N_T$ different temperatures.
Here, we need two copies of the system at each temperature
to calculate $q$, so we actually
run 2 sets of $N_T$ replicas.
We also gain a large speed-up by
using multispin coding\cite{multispin}
to store each spin or bond as a single bit rather than
a whole word.

In earlier work\cite{kpy} for the Gaussian
distribution we were able to use a special relationship between certain
variables to check for equilibration, but this is not applicable here.
We therefore
investigate whether various quantities have become independent of simulation
time when plotted on a logarithmic scale. Fig.~\ref{equil} shows an example
for $L=8, T =0.20$ indicating that the data seems to have saturated.

\begin{table}
\begin{center}
\begin{tabular}{lrrrr}
L &  $\nsa$      & $\nsw$          & $N_T$ & $T_{min}$ \\ 
\hline 
4 & 9600         &   $10^5$        & 15 & 0.05\\
6 & 6400         &   $10^6$        & 15 & 0.05 \\
8 & 3904$^{(*)}$ & 3 $\times 10^6$ & 21 & 0.2  \\
10 & 1408        &   $ 10^7$       & 19 & 0.35 
\end{tabular}
\end{center}
\caption{
Parameters of the simulations. $\nsa$ is the
number of samples (i.e. sets of bonds), $\nsw$ is the total number of sweeps
simulated for each of the $2 N_T$ replicas for a single sample,
$N_T$ is the number of
temperatures used in the parallel tempering method, and
$T_{min}$ is the lowest temperature simulated.  $^{(*)}$  $\nsa$=6336 
for $L=8$ and $T\ge 0.35$.}
\label{3d-tab}
\end{table}

In Table~\ref{3d-tab}, we show the simulation parameters.
The lowest temperature simulated, $T_{min}$, has 
to be compared with\cite{tc} $T_c \approx 1.15$. For each size the 
largest temperature is 2.0. 
The set of temperatures is determined by requiring that 
the acceptance ratio for global moves is 0.3 or larger.

\begin{figure}
\begin{center}
\epsfxsize=\columnwidth
\epsfbox{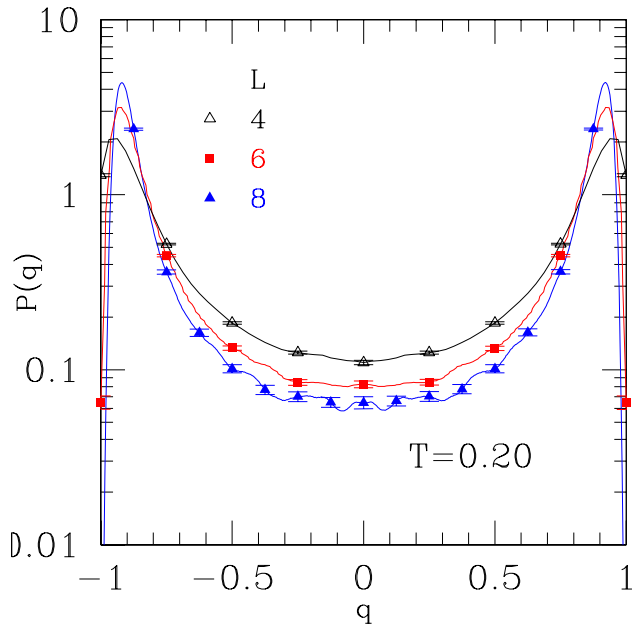}
\end{center}
\caption{Data for the overlap distribution $P(q)$ at $T=0.20$. The
vertical scale is logarithmic to better make visible the peak at large
$q$ and the tail down to $q=0$. 
We only display {\em some}\/ of the data points
as symbols, for clarity, but the lines connect {\em all}\/ the data points.
This accounts for the curvature between neighboring
symbols.
}
\label{pq0.2}
\end{figure}

\begin{figure}
\begin{center}
\epsfxsize=\columnwidth
\epsfbox{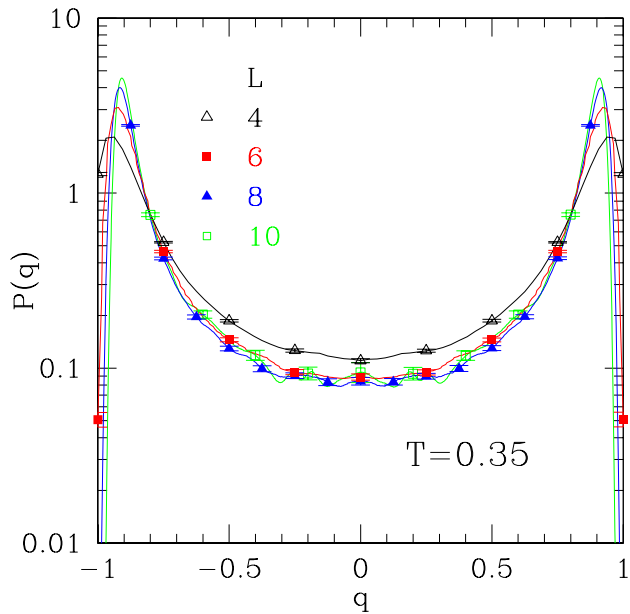}
\end{center}
\caption{Same as for Fig.~\ref{pq0.2} but at $T=0.35$.
}
\label{pq0.35}
\end{figure}

\begin{figure}
\begin{center}
\epsfxsize=\columnwidth
\epsfbox{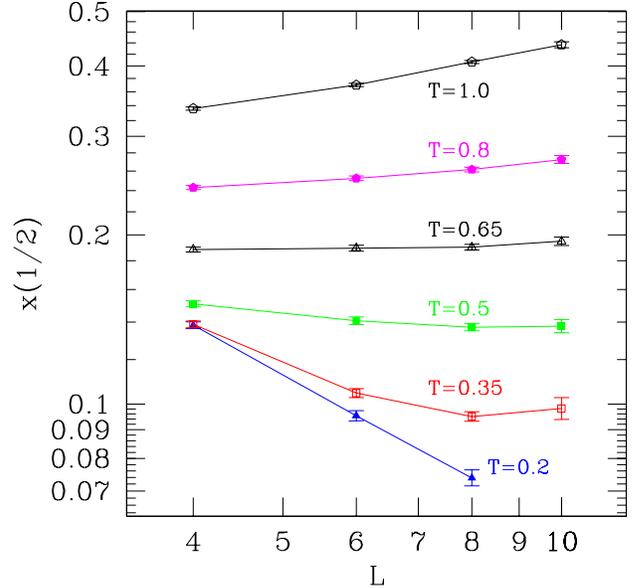}
\end{center}
\caption{Log-log plot of $x(1/2)$, the average of $P(q)$ over $|q|\leq 1/2$,
against $L$.
}
\label{p0}
\end{figure}

Figs.~\ref{pq0.2} and \ref{pq0.35} show data for $P(q)$ for different sizes 
at $T=0.2$ and $T=0.35$. One
can see that the weight in the tail tends to decrease initially
with increasing $L$, especially at lower $T$, but for $T=0.35$ the data
seems to saturate at larger $L$. For $T=0.5$ (not displayed) the weight
in the tail saturates already at $L=4$.
This can be seen more clearly in Fig.~\ref{p0}, which shows $x(1/2)$
as a function of $L$ for different temperatures, where
$x(q)=\int_{-q}^{q} P(q') dq'$ so $x(1/2)$ is the {\em average}\/
of $P(q)$ from $-1/2$ to $1/2$. We give data for $x(1/2)$ rather than $P(0)$
because the statistics are better and also so we can compare directly with other
work.
For $L=4$ and $L=6$, the data at $T=0.05$, not showed in Fig.~\ref{p0}, are 
superimposed to the data at $T=0.2$, indicating that we have reached the
true $T=0$ behavior. For $L=8$, the data at $T=0.2$ may be
one or two standard deviations larger than the $T=0$ value.
Furthermore, the average energy at $T=0.2$ agrees 
with the ground state results by Pal\cite{pal}. From a power law fit of the 
data in Fig.~\ref{p0} at $T=0.2$ we estimate
\begin{equation}
\lambda=0.9 \pm 0.1 \, .
\label{lambda}
\end{equation}

Generalizing the KM argument to a scenario described by an exponent
$\theta$, we expect that at finite-$T$  there will be a crossover 
between the $L^{-(\lambda+\theta)}$ behavior for $L$ smaller than some
length scale $L_c(T)$, and the $L^{-\theta}$ behavior
(or, for $\theta=0$, an $L$-independent value proportional 
to $T$), at scales larger than $L_c(T)$. 
In the more general case, assuming scaling one has $L_c(T) \sim T^{-1/\lambda}$ 
and 
\begin{equation}
x(1/2) = T L^{-\theta} f(L T^{1/\lambda}) ,
\label{scaling_eq}
\end{equation}
where $f$ is a scaling function.

\begin{figure}
\begin{center}
\epsfxsize=\columnwidth
\epsfbox{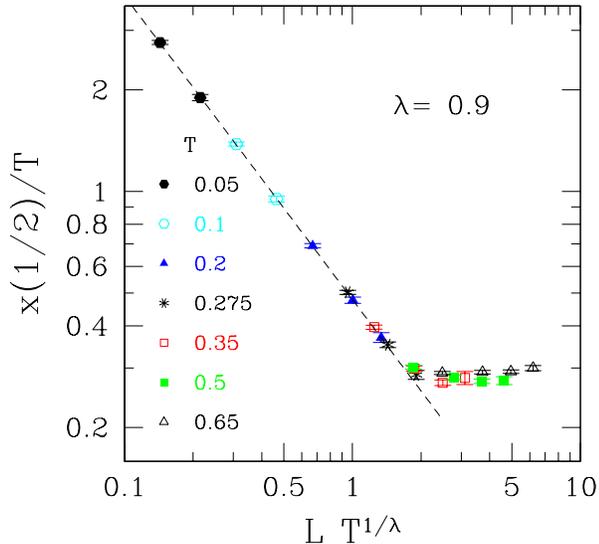}
\end{center}
\caption{
The scaling behavior of $x(1/2)$ expected from
Eq.~(\ref{scaling_eq}) with $\theta=0$. 
For $L \gg L_c(T) \sim T^{-1/\lambda}$, $x(1/2)$ is
independent of size, while for $L \ll  L_c(T), x(1/2)$ varies as 
$L^{-\lambda}$. The dashed line has a slope of $-0.9$.
}
\label{scaling}
\end{figure}

A scaling plot appropriate to this behavior, for $\theta=0$ and 
 $\lambda=0.9$, is shown in Fig.~\ref{scaling}, where
one can see that the data collapse fairly well.
The data in Fig.~\ref{p0} {\em increase}\/ with increasing
$L$ for $T\ge 0.8$,
due to the vicinity of $T_c$, where
$x(1/2) \sim L^{\beta/\nu}$ and\cite{tc} $\beta/\nu\simeq 0.3$.
One may therefore argue\cite{moore} that the observed saturation 
between $T=0.35$ and $T=0.65$ 
is a finite size effect and that at larger sizes there will be a 
{\em second}\/ crossover to the $L^{-\theta}$ behavior. We cannot exclude this
possibility,
though we note that $T=0.35$ is quite far from $T_c$ and that a scaling 
plot as in  Fig.~\ref{scaling} but with $\theta=0.2$ is significantly 
worse.

Hartmann\cite{hart2} computed $x(1/2)$ as a function of 
$L$ at zero temperature and found that a power law fits well the data 
with an exponent $\lambda = 1.25 \pm 0.05$, which disagrees with our estimate.
Our value for $\lambda$ does, however,
agree with that of Berg et al.\cite{berg} who find
$\lambda = 0.72\pm 0.12$. In addition, our raw data for $x(1/2)$ is consistent 
with (though more accurate than) that of Berg et al.\cite{berg}, but is 
inconsistent with that of  Hartmann\cite{hart2} for $L>4$. 
For example, for $L=6$ we find $x(1/2)=0.095 \pm 0.002$, while Hartmann 
finds  $x(1/2)= 0.083 \pm 0.005$. We note however that 
Hartmann's method, unlike (properly equilibrated) Monte Carlo simulations, 
is not {\em guaranteed}\/ to sample all the ground states with equal probability.

Our results for $P(q)$ at low-T are also in marked disagreement with HG. 
For example, HG report a $P(q)$ which is lower
than $0.03$ in the interval $|q|\le 0.1$ for $L=8$ and $T=0.3$, while 
our average of $P(q)$ over this interval is between $0.066 \pm 0.004$ 
(our value at $T=0.275$) and $0.081 \pm 0.004$ (our value $T=0.35$).
HG observe a pronounced decrease of $P(q)$ with $L$ even at $T=0.5$, where our 
data clearly saturate. We also computed the Binder cumulant, which agrees with
Ref.~\onlinecite{mprz} but disagrees with HG. This suggests
that the simulations 
of HG are not correctly equilibrated, as discussed in detail in 
Ref.~\onlinecite{mprz}.

KM give arguments that $\lambda$ should equal $d_s/2$ where $d_s$ is the
fractal dimension of the surface of the large-scale low-energy excitations
which give rise to a nontrivial energy landscape. However, one expects
that \mbox{$d_s \ge d-1$} which is barely satisfied by the estimate in
Eq.~(\ref{lambda}) which corresponds to $d_s = 1.8 \pm 0.2$.
Furthermore, for the Gaussian distribution, $d_s$ is
significantly larger than this value. For example, Ref.~\onlinecite{py} finds
$d_s = 2.58 \pm 0.02 $. While it is possible that $d_s$ could be
different for the Gaussian and $\pm J$ models, our results suggest
that $\lambda \ne d_s/2$, and that there may be corrections to
the argument of KM.

To conclude, results from simulations on small sizes indicate that the order
parameter distribution of the $\pm J$ Ising spin glass is trivial at $T=0$
but, at least for quite small sizes, is
nontrivial at finite-$T$ in agreement with the conclusions of KM. We have also
demonstrated crossover scaling
between the zero-$T$ and finite-$T$ behaviors. We expect similar results in
other models with a discrete disorder distribution, and indeed this is what we
find in preliminary unpublished data for the $\pm J$ Ising spin glass in
$d=4$.  Whether these conclusions are still valid in the thermodynamic limit
remains an open question.  However, we emphasize, quite generally, that a
trivial $P(q)$ at $T=0$ does {\em not}\/, in itself, imply evidence for the
droplet model since this is also expected if $P(q)$ is nontrivial at
finite-$T$, as pointed out by KM.

After this work was submitted we received a paper by Hed et al.\cite{hed}, in
which, based on a different analysis from ours, they claim that $P(q)$ is
non-trivial at $T=0$.

This work was supported by the National Science Foundation under grant DMR
0086287.  The numerical calculations were made possible by a grant of time
from the National Partnership for Advanced Computational Infrastructure.
We should also like to thank A.~Hartmann for helpful correspondence.


\begin{references}
\bibitem{extrap}
    We make no comments here as to whether these conclusions will still be
    valid in the thermodynamic limit.
\end{references}
\end{document}